\documentclass[aos,preprint]{imsart}

\RequirePackage[OT1]{fontenc}
\RequirePackage{amsthm,amsmath}
\RequirePackage[numbers]{natbib}
\RequirePackage[colorlinks,citecolor=blue,urlcolor=blue]{hyperref}

\startlocaldefs
\numberwithin{equation}{section}
\theoremstyle{plain}
\newtheorem{thm}{Theorem}[section]
\endlocaldefs

\usepackage{amsfonts}
\usepackage{mathrsfs}
\usepackage{amsthm}
\usepackage{multirow}
\usepackage{bbding}
\usepackage{amssymb}
\usepackage{amsmath}
\usepackage{graphicx,color}
\usepackage{rotating}
\usepackage{bm}

\DeclareMathOperator*{\argmin}{arg\,min}

\usepackage{float}

\makeatletter

\newcommand{\Rmnum}[1]{\expandafter\@slowromancap\romannumeral #1@}
\makeatother

\begin{document}

\begin{frontmatter}
\title{A simple and efficient profile likelihood for semiparametric exponential family\thanksref{T1}}
\runtitle{profile likelihood for semiparametric exponential family}
\thankstext{T1}{The research was supported by NNSF projects (11571204 and 11231005) of China.}

\begin{aug}
\author{\fnms{Lu} \snm{Lin}\thanksref{t1,t2}\ead[label=e1]{linlu@sdu.edu.cn.}},
\author{\fnms{Lili} \snm{Liu}\thanksref{t2}}
\and
\author{\fnms{Xia} \snm{Cui}\thanksref{t3}}

\thankstext{t1}{The corresponding author}
\runauthor{Lu Lin, Lili Liu and Xia Cui}
\affiliation{Shandong University\thanksmark{t2} and Guangzhou University\thanksmark{t3}}

\address{Lu Lin and Lili Liu\\Zhongtai Securities Institute for Financial Studies\\Shandong University\\ China\\
\printead{e1}\\
\phantom{E-mail:\ }}

\address{Xia Cui\\School of Economics and Statistics\\Guangzhou University\\China\\}
\end{aug}

\begin{abstract}

Semiparametric exponential family proposed by Ning et al. (2017) is an extension of the parametric exponential family to the case with a nonparametric base measure function. Such a distribution family has potential application in some areas such as high dimensional data analysis. However,
the methodology for achieving the semiparametric efficiency has not been proposed in the existing literature. In this paper, we propose a profile likelihood to efficiently estimate both parameter and nonparametric function. Due to the use of the least favorable curve in the procedure of profile likelihood, the semiparametric efficiency is achieved successfully and the estimation bias is reduced significantly. Moreover, by making the most of the structure information of the semiparametric exponential family, the estimator of the least favorable curve has an explicit expression. It ensures that the newly proposed profile likelihood can be implemented and is computationally simple. Simulation studies can illustrate that our proposal is much better than the existing methodology for most cases under study, and is robust to the different model conditions.

\end{abstract}



\end{frontmatter}

\section{Introduction}

\subsection{Parametric exponential family}
Exponential family may be the most popular distribution pattern. This special form is chosen for mathematical convenience, due to some useful algebraic properties, as well as for generality, because it is a very natural set of distributions in some sense. The classical exponential family contains common and important distributions such as normal distribution, exponential distribution, gamma distribution, chi-squared distribution, beta distribution and so on.
Specifically, a random variable $Y\in\mathscr Y\subseteq \mathbb{R}$ follows the natural exponential family with unknown parameter $\theta$, if its density has the form:
\begin{equation}\label{(0.1)} p(y;\theta)=\exp\left\{\theta\cdot y-b(\theta)+\log f(y)\right\}.\end{equation}
In the distribution family, $f(\cdot)$ is a given base measure function and then $b(\cdot)$ is a known function. Moreover, if a covariate vector $X\in {\mathscr X}\subset \mathbb{R}^d$ is available, the random variable $Y$ follows given $X$ the natural exponential family with unknown parameter $\beta$ and canonical link
$\theta({x})=\beta^Tx$, if its density can be expressed as
\begin{equation}\label{(0.2)}p(y;\beta|{x})=\exp\left\{\theta({x})\cdot y-b(\theta({x}))+\log f(y)\right\}.\end{equation} Here the regression coefficient $\beta$ characterizes the covariate effect, and the known function $f(\cdot)$ specifies a certain distribution in the natural exponential family.
For instance, the choice of $f(y)=\exp\{-y^2/2\}$ results in a linear regression with standard Gaussian noise; the choice of $f(y)=1$ for $y=0,1$ leads to a logistic regression; and the choice of $f(y)=1/y!$ for $y=0,1,2,\cdots$ induces a Poisson regression.

\subsection{Semiparametric exponential family}

\subsubsection{Semiparametric framework}

It is worth pointing out that
in the classical exponential families (\ref{(0.1)}) and (\ref{(0.2)}), the base measure function $f(\cdot)$ is always supposed to be known. A natural problem now is how to extend the distribution family to the case with unknown function $f(\cdot)$. Recently, Ning et al. (2017) provided some examples to show that under some situations the parametric exponential family distributions can result in a semiparametric exponential family distribution with an unknown base measure function, and such a semiparametric exponential family has potential application in high dimensional data analysis, including incomplete data, selection bias and heterogeneity.

Specifically, a random variable $Y\in\mathscr Y\subseteq \mathbb{R}$ satisfies a semiparametric natural exponential family (spEF) with unknown parameters $(\theta,f)$, if its density has the form:
\begin{equation}\label{(1.1)}p(y;\theta,f)=\exp\left\{\theta\cdot y-b(\theta,f)+\log f(y)\right\},\end{equation}
where $f(\cdot)$ is an unknown base measure function, $\theta$ is an unknown canonical parameter and $b(\theta,f)=\log\int_{\mathscr Y}\exp\{\theta\cdot y\}f(y)dy<\infty$ is the log-partition function. More conditions and explanations for the model will be given in the next section.

Moreover, for a $d$-dimensional covariate $X\in {\mathscr X}\subset \mathbb{R}^d$, suppose the canonical linear link $\theta=\beta^Tx$. Then, $Y$ given $X$ follows the spEF with unknown parameters $(\beta,f)$ if its density can be expressed as
\begin{equation}\label{(1.2)}p(y;\beta,f|{x})=\exp\left\{\beta^T{x}\cdot y-b(\beta^T{x},f)+\log f(y)\right\}.\end{equation}
In the spEF, the newly defined base measure function $f(y)$ can be thought of as an infinite dimensional parameter. With suitable choices $f(y)$, the spEF recovers the whole class of natural exponential
family distributions, showing the spEF extends the classical natural exponential family to the case with infinite dimensional base measure $f(\cdot)$.

\subsubsection{Existing methodology and related concerns}

The existing method of Ning et al. (2017) focuses only on the parametric component $\beta$ but regards the nonparametric component $f(\cdot)$ as a nuisance parameter. By a conditional likelihood, the nonparametric component $f(\cdot)$ is eliminated and then the parametric component $\beta$ can be estimated consistently. Suppose that $(X_i,Y_i),i=1,\cdots,n$, are independent observations from (\ref{(1.2)}). Denote ${\bf Y}=(Y_1,\cdots,Y_n)$ and ${\bf X}=(X_1,\cdots,X_n)$, and let ${\bf R}=(R_1,\cdots,R_n)$ and ${\bf Y}_{(\cdot)}=(Y_{(1)},\cdots,Y_{(n)})$ be the rank and order statistic of $\bf Y$, respectively. Then, given ${\bf X}$, the conditional density $p({\bf y};\beta,f|{\bf x})$ of $\bf Y$ can be expressed
\begin{equation}\label{(1.8)}p({\bf y};\beta,f|{\bf x})=P({\bf R}={\bf r};\beta,f\,|\,{\bf x},{\bf y}_{(\cdot)})\cdot p({\bf y}_{(\cdot)};\beta,f\,|\,{\bf x}).\end{equation} As shown by Ning et al. (2017), the first factor on the right hand side of the equation (\ref{(1.8)}) is free of $f$, i.e., $P({\bf R}={\bf r};\beta,f\,|\,{\bf x},{\bf y}_{(\cdot)})=P({\bf R}={\bf r};\beta\,|\,{\bf x},{\bf y}_{(\cdot)})$.
Without the nonparametric function $f(\cdot)$, Ning et al. (2017) use the conditional likelihood \begin{equation}\label{(1.9)}l(\beta\,|\,{\bf y}_{(\cdot)})=P({\bf R}={\bf r};\beta\,|\,{\bf x},{\bf y}_{(\cdot)})\end{equation} to infer the parameter of interesting $\beta$. The estimation consistency can be achieved successfully.

When looking at the conditional likelihood (\ref{(1.9)}), however, we have the following concerns:
\begin{itemize}\item[(1)] {\it Loss of semiparametric efficiency.} By comparing the conditional likelihood (\ref{(1.9)}) with the full likelihood (\ref{(1.8)}), we can see that the conditional likelihood (\ref{(1.9)}) does not contain full information of $\beta$. More specifically, the second factor $p({\bf y}_{(\cdot)};\beta,f\,|\,{\bf x})$ on the right hand side of (\ref{(1.8)}), which is excluded from the conditional likelihood, also contains the information of the parameter of interesting $\beta$. Consequently, the estimation efficiency will be lost only by conditional likelihood (\ref{(1.9)}).
\item[(2)] {\it Computational complexity and estimation bias.} Moreover, the conditional likelihood (\ref{(1.9)}) is computationally intensive due to the
combinatorial nature of permutations. To this end,
Ning et al. (2017) employed the $2$th order information form of the conditional likelihood (\ref{(1.9)}) as its surrogate, which has the form:
   \begin{equation}\label{(1.11)}l(\beta)= -\left(\begin{array}{cc}n\\2\end{array}\right)^{-1}\sum_{1\leq i< j \leq n}\log\left(1+R_{ij}(\beta)\right),\end{equation} where $R_{ij}(\beta)=\exp\{-(Y_i-Y_j)\cdot\beta(X_i-X_j)\}$.
For easy examining the behavior of $l(\beta)$, we consider the univariate regression as
\begin{equation}\label{(1.10)}Y_i=\beta X_i+\varepsilon_i,\ i=1,\cdots,n,\end{equation} where the scalar parameter $\beta>0$, the errors $\varepsilon_1,\cdots,\varepsilon_n$ are independent and identically distributed as an exponential family distribution with known $\sigma^2$. It can be seen from the regression (\ref{(1.10)}) that if $\sigma^2$ is small, then
  $$(Y_i-Y_j)\cdot\beta(X_i-X_j)>0 \mbox{ for most } i\in\{1,\cdots,n\}.$$
The above indicates that to maximize the conditional likelihood function $l(\beta)$ in (\ref{(1.11)}), the positive parameter $\beta$ should has a relatively large value, instead of the true value of $\beta$. Thus, it is quite possible that the conditional likelihood in (\ref{(1.11)}) is not maximized at the true value of $\beta$ for the case of small variance $\sigma^2$.
Under this situation, the resulting conditional likelihood estimator may have a large bias.
\end{itemize}

For illustrating the issue of estimation bias aforementioned,
we now try a simple simulation. Under regression (\ref{(1.10)}), let $X_i,i=1,\cdots,n$, be independent and identically distributed as $N(0,1)$, and $\varepsilon_i,i=1,\cdots,n$, be independent and identically distributed as $N(0,\sigma^2)$. With the different choices of $\sigma^2=0.05$, 0.1 and 1, Figure 1 presents the estimated curves of the median of the conditional likelihood $l(\beta)$ defined in (\ref{(1.11)})
for the case of $\beta=2$ and $n=100$, through 100 replications. It can be clearly seen that for the small choices of $\sigma^2=0.05$ and 0.1, the estimated curves of $l(\beta)$ are almost overlapped, and are increasing functions of $\beta$ in the interval $[0,10]$. In this case, the maximum value of the conditional likelihood function $l(\beta)$, if it exists, should be achieved at a value of $\beta$ beyond 10, instead of the true value $\beta=2$. This illustrates that the resulting conditional likelihood estimator of $\beta$ has a huge bias for the case of small $\sigma$. This point of view will be further illustrated by more simulation studies given in Section 4.

\begin{figure}[H]
\caption[]{The estimated curves of median of $l(\beta)$ for $\sigma^2=0.05$ (dashed line), $\sigma^2=0.1$ (solid line) and $\sigma^2=1$ (dotted line).}
\centering
\includegraphics[height=7cm]{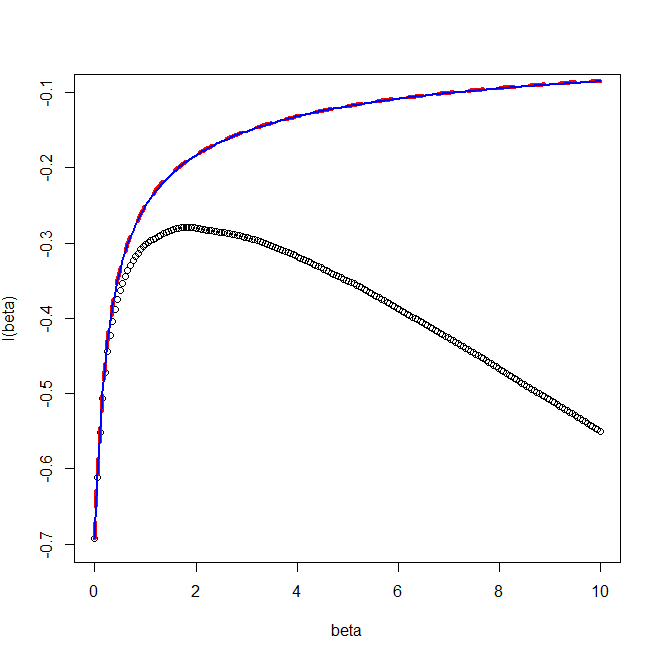}
\end{figure}

\subsection{Profile likelihood and semiparametric efficiency}

For the model with unknown $f(\cdot)$, how to estimate both $\beta$ and $f(\cdot)$, and how to achieve semiparametric efficiency are significant issues in the procedure of statistical inference.
It is clear that full likelihood (or profile likelihood), instead of conditional likelihood, contains the complete information of the parameter and nonparametric function in the model. The utilization of the full likelihood (or profile likelihood) can help improving statistical analysis and achieving estimation efficiency.

Semiparamtric efficiency has always been an important issue in the area of semiparametric statistics. It is known that profile likelihood and the extended versions are effective methodologies for achieving semiparametric efficiency. Severini and Wang (1992) introduced a profile likelihood for conditionally parametric model, a special semiparametric model that contains parameter and nonparametric function. Severini and Staniswalis (1994) proposed a quasi-likelihood, a special profile likelihood, for a semiparametric model that is defined by the conditions of the first two moments. Lin et al. (2005) suggested a profile empirical likelihood for a semiparametric estimating equation. General and comprehensive theories of semiparametric models and semiparametric efficiency can be found in Bikckel et al. (1993).
Furthermore, the idea of semiparametric efficiency has been wildly applied in various models such as partially linear model (see, e.g., Engle et al. (1986) and H\"{a}rdle et al. (2000)), single-index model and its extended versions (see, e.g., McCullagh and Nelder (1989),
Hristache et al. (2001), and Zhu and Xue (2006)), and the regression with missing data (Robins and Rotnitzky (1995)) or with errors-in-variables (Liang et al. (1999)).

\subsection{Our contributions}

In this paper, we propose a semiparmatric profile likelihood for efficiently estimating both parametric and nonparametric components in the semiparametric exponential family. This paper contains three major contributions.
\begin{itemize}\item[(1)] Due to the use of the least favorable curve in the procedure of profile likelihood, the semiparametric efficiency is successfully achieved and the estimation bias is significantly reduced.
\item[(2)] By the relation between the expectation of $Y$ and the derivative of function $b(\theta,f)$, the least favorable curve obtains an explicit expression. Consequently, the the nonparametric estimator of $f(y)$ can be easily estimated, and then the whole procedure of the profile likelihood can be easily implemented.
    \item[(3)] Our methodology can be applied to some commonly used models such as regressions with missing data or heterogeneity, and can be used for other statistical inferences such as the constructions of confidence interval and hypothesis test.
\end{itemize}

The remainder of the paper is organized in the following way. In Section 2, the semiparametric exponential family is briefly recalled, a motivating example is introduced and some related problems are discussed. In Section 3, the semiparametric profile likelihood for the semiparametric exponential family is introduced, its estimation procedure is suggested and theoretical property is investigated.
In Section 4, the finite sample properties of the proposed procedures are evaluated via simulation studies.

\setcounter{equation}{0}
\section{Model recognition and estimating equation}

\subsection{Model recognition and example} Before introducing the profile likelihood for the semiparametric exponential family, we further recognize the spEFs defined by (\ref{(1.1)}) and (\ref{(1.2)}). The constraint $\int_{\mathscr Y}\exp\{\theta\cdot y\}f(y)dy<\infty$ in the definitions means that the unknown base measure function $f(\cdot)$ is not heavy-tailed. Note that spEF$(\theta,f)$ is identical to spEF$(\theta,cf)$ for any constant $c\neq0$. To address this problem, we need to impose the identifiability condition:
$$\int_{\mathscr Y}f(y)dy=1.$$ On the other hand, some exponential family distributions, such as the normal distribution, involve dispersion parameters. Under this situation, we can re-parameterize the parameters so that the resulting model can be expressed as a semiparametric natural exponential family with newly defined parameters. For the details see Ning et al. (2017).

To further illustrate the model and motivate our study, we look at the following simple model. Suppose that $Y$ given $X=x$ follows the normal distribution $N(\theta(x),1)$ with $\theta(x)=\beta^Tx$. In this case, the density function of $Y$ given $X$ has the form:
\begin{equation}\label{(1.4)}p(y;\beta|{x})=\exp\left\{\beta^T{x}\cdot y-\frac{1}{2}(\beta^Tx)^2+\log \frac{1}{\sqrt{2\pi}}\exp\left(-\frac{y^2}{2}\right)\right\}.\end{equation}
Consider the case when the sample
$\{(Y_i,X_i),i=1,\cdots,n\}$ of $(Y,X)$ contains missing values. Define an indicator variable $\delta_i$ by that $\delta_i=1$ if $(Y_i,X_i)$ is observed, and $\delta_i=0$ otherwise. Suppose the decomposable framework as $P(\delta_i=1|Y_i,X_i)=g_1(Y_i)g_2(X_i)$ for some unknown functions $g_1(\cdot)$ and $g_2(\cdot)$, satisfying $\int_{\mathscr Y}g_1(y)dy=1$ or $\int_{\mathscr X}g_2(x)dx=1$. As shown by Ning et al. (2017), such a decomposable assumption is common, and for the observed data, the following distribution holds:
\begin{equation}\label{(1.5)}p(Y_i;\beta|X_i,\delta_i=1)=\exp\left\{\beta^TX_i\cdot Y_i-b(\beta^TX_i,f^m)+\log f^m(Y_i)\right\},\end{equation} where $f^m(y)=\frac{1}{\sqrt{2\pi}}\exp(-\frac{y^2}{2})g_1(y)$. Similar to the spEF in (\ref{(1.2)}), the distribution above contains an unknown base measure function $f^m(y)$ because of the use of the unknown function $g_1(y)$. By the definition of $b(\cdot,\cdot)$, we have
\begin{eqnarray}\label{(1.6)}\nonumber b(\beta^TX_i,f^m)&=&\log\int_{\mathscr Y}\exp\{\beta^TX_i\cdot y\}f^m(y)dy\\&=&\frac12(\beta^TX_i)^2+
\log\int_{\mathscr Y}\frac{1}{\sqrt{2\pi}}\exp\left\{-\frac{(y-\beta^TX_i)^2}{2}\right\}g_1(y)dy.\end{eqnarray}
It can be seen that $$E(Y_i|X_i,\delta_i=1)=b'(\beta^TX_i,f^m),$$ where $b'(\beta^TX_i,f^m)$ is the derivative of $b(\beta^TX_i,f^m)$ with respect to $\theta=\beta^TX_i$. With the observed data $(Y_i,X_i,\delta_i=1)$, we then get the outcome regression model as
\begin{equation}\label{(1.7)}E(Y_i|X_i,\delta_i=1)=\beta^TX_i+g(\beta^TX_i,g_1),\end{equation} where $$g(\beta^TX_i,g_1)=\frac{\int_{\mathscr Y}(y-\beta^TX_i)\exp\left\{-\frac{(y-\beta^TX_i)^2}{2}\right\}g_1(y)dy}{\int_{\mathscr Y}\exp\left\{-\frac{(y-\beta^TX_i)^2}{2}\right\}g_1(y)dy}.$$

Here $g(\cdot,g_1)$ is unknown or a known functional of the unknown function $g_1$. Thus, the resulting regression (\ref{(1.7)}) is actually a generalized partially linear model or generalized
partially linear single-index model with linear part $\beta^TX_i$ and nonparametric component $g_1$ or $g(\cdot,g_1)$. The use of $g(\cdot,g_1)$, instead of ignoring it, can improve the estimation precision; see for example
Carroll et al. (1997), Severini and Wang (1992) and Severini and Staniswalis (1994).

\subsection{Estimating equation}

It will be proved in Appendix that for the spEF in (\ref{(1.2)}), the corresponding score functions have zero expectations:
\begin{equation}\label{(1.3)}E\left(\frac{\partial l(\beta,f)}{\partial \beta}\right)=0 \ \mbox{ and } E\left(\frac{\partial l(\beta,f)}{\partial f}\Big|y\right)=0,\end{equation} where $l(\beta,f)=\beta^T x\cdot y-\log \int_{\mathscr Y}\exp\{\beta^T x\cdot y\}f(y)dy+\log f(y)$, the log-likelihood function of $\beta$ and $f$, $\frac{\partial l(\beta,f)}{\partial \beta}$ is the derivative of function $l(\beta,f)$ with respect to parameter $\beta$, and $\frac{\partial l(\beta,f)}{\partial f}$ is the functional derivative of $l(\beta,f)$ with respect to function $f$.
The above indicates that theoretically the true values of $\beta$ and $f$ could be defined as the solutions of the equations in (\ref{(1.3)}).
Thus, both equations in (\ref{(1.3)}) can be used in the procedure of constructing the efficient estimators.
The above is the key for our methodological development. The remaining tasks are how to solve the equations (including the functional equation) and how to construct the related estimators.

The main difference from the classical semiparametric models such as partially linear regression is that in the new semiparametric exponential families (\ref{(1.1)}) and (\ref{(1.2)}), and the equation (\ref{(1.3)}), $b(\beta,f)$ is a functional of unknown function $f$, rather than a simple function. Thus, we need special techniques to deal with the issue. Based on the relation between the conditional expectation of $Y$ and the derivative of function $b(\theta,f)$, we use the conditional expectation to replace the functional $b(\beta,f)$ and then use nonparametric kernel function to estimate the related unknowns. For the details see the next section.

\setcounter{equation}{0}
\section{Profile likelihood for semiparametric exponential family}

All the issues, including the estimation efficiency loss and the estimation bias increase caused by the conditional likelihood aforementioned in Introduction, and the new outcome regression (\ref{(1.7)}) and the new estimating equation (\ref{(1.3)}) given in Section 2, motivate us to explore novel approach.
In this section, we propose a profile likelihood for estimating both the parametric component $\beta$ and nonparametric component $f(\cdot)$, efficiently.

\subsection{Least favorable curve}
To achieve semiparametric efficiency, by the profile likelihood theory in semiparametric models (see Severini and Wang (1992)), we need to find a least favorable curve, a function: $\beta\in\mathbb B\mapsto f_\beta\in\mathbb{R}$, with which the estimation efficiency for parameter $\beta$ can be achieved. It can be seen that the functional derivative of $l(\beta,f)$ with respect to $f$ has the form:
\begin{equation}\label{(3.1)}-\frac{\exp\{\beta^T x\cdot y\}}{ \int_{\mathscr Y}\exp\{\beta^T x\cdot y\}f(y)dy}+\frac{1}{f(y)}.\end{equation} The proof for (\ref{(3.1)}) is given in Appendix. By this and the second equation in (\ref{(1.3)}), the function $f(y)$ should satisfy the equation $$E\left(-\frac{\exp\{\beta^T X\cdot y\}}{ \int_{\mathscr Y}\exp\{\beta^T X\cdot y\}f(y)dy}+\frac{1}{f(y)}\Big|y\right)=0,$$ which can be further expressed as
\begin{equation}\label{(3.2)}E\left(-\frac{\exp\{\beta^T X\cdot y\}}{ \exp\{b(\beta^T X, f)\}}+\frac{1}{f(y)}\Big|y\right)=0.\end{equation}

On the other hand, like the expectation property of the classical natural exponential family, the conditional expectation $E(Y|{x})$ and the function $b(\beta{x},f)$ in the spEF defined by (\ref{(1.2)}) have the following relation:
\begin{equation}\label{(3.3)}b'(\beta^T{x},f)=E(Y|{x})=E(Y|\beta^T x),\end{equation} where $b'(\beta^T{x},f)$ is the derivative of $b(\beta^T{x},f)$ with respect to $\theta=\beta^T{x}$. By the equation above, the definition of $b(\beta^T{x},f)$ and the identifiability condition
$\int_{\mathscr Y}f(y)dy=1$, we have
\begin{eqnarray}\label{(3.4)}b(\beta^T{x},f)&=&\int_{0}^{\beta^T x}b'(\beta^T{x},f)d(\beta^T x)+b(0,f)\\&=&\int_{0}^{\beta^T x}E(Y|\beta^T x)d(\beta^T x).\nonumber \end{eqnarray} Substituting this into equation (\ref{(3.2)}), we then get the solution of $f(y)$ as
\begin{equation}\label{(3.5)}f_\beta(y)=\left(E\left(\frac{\exp\{\beta^T X\cdot y\}}{ \exp\{\int_{0}^{\beta^TX}E(Y|\beta^T x)d(\beta^T x)\}}\Big|y\right)\right)^{-1}.\end{equation}

It will be verified in the proof of Theorem 1 given below that actually $f_\beta(y)$ is a least favorable curve. Moreover, when $\beta=\beta^0$, the true value of $\beta$, the least favorable curve is equal to the true function $f(y)$, namely,
$$f_{\beta^0}(y)=f(y).$$
The use of such a least favorable curve is the key for achieving the semiparametric efficiency.

\subsection{Estimation methodology}

Here we only consider the case when $y$ is a continuous variable and $f(y)$ is a smooth function of $y$. For the case of discrete variable, we can use
the discrete kernel, together with the method proposed below, to build the nonparametric estimation (see, e.g., Aitchison and Aitken (1976), Racine and Li (2004), Li and Racine (2007), and Chen and Tang (2011)), but the details are omitted here.

Let $\{(Y_i,X_i):i=1,\cdots,n\}$ be a sample of $(Y,X)$. For given $\beta$, the conditional expectation $E(Y|\beta^T x)$ can be estimated by common nonparametric methods. For example, its N-W estimator has the form:
$$\widehat E(Y|\beta^T x)=\frac{\sum_{i=1}^n Y_iK(\beta^T(X_i-x)/h)}{\sum_{i=1}^n K(\beta^T(X_i-x)/h)},$$ where $K(\cdot)$ is a kernel function and $h$ is the bandwidth depending on $n$.
We then estimate $\exp\{ \int_{0}^{{\beta^T X_i}}E(Y|\beta^Tx)d(\beta^Tx)\}$
by
$$\exp\left\{\int_0^{\beta^TX_i}\frac{\sum_{j=1}^n Y_jK(\beta^T (X_j-t)/h)}{\sum_{j=1}^n K(\beta^T (X_j-t)/h)}dt\right\}.$$ Similarly, we can use a kernel method to estimate the conditional expectation $E\left(\cdot|y\right)$. By substituting these into the equation (\ref{(3.5)}), the estimators of $f_\beta(y)$ and $f_\beta(Y_k)$ are obtained as
\begin{eqnarray}\label{(3.6)} &&\widetilde f_\beta(y)=\left(\sum_{i=1}^n\frac{\exp\{\beta^T X_i\cdot y\}}{ \exp\left\{\int_0^{\beta^TX_i}\frac{\sum_{j=1}^n Y_jK((\beta^T X_j-t)/h)}{\sum_{j=1}^n K((\beta^T X_j-t)/h)}dt\right\}}W_i(y)\right)^{-1},\\\label{(3.7)}&&
\widetilde f_\beta(Y_k)=\left(\sum_{i\neq k,i=1}^n\frac{\exp\{\beta^T X_i\cdot Y_k\}}{ \exp\left\{\int_0^{\beta^TX_i}\frac{\sum_{j=1}^n Y_jK((\beta^T X_j-t)/h)}{\sum_{j=1}^n K((\beta^T X_j-t)/h)}dt\right\}}W_i(Y_k)\right)^{-1}.\end{eqnarray} where the weights $$W_i(y)=\frac{ K((Y_i-y)/h)}{\sum_{i=1}^n K((Y_i-y)/h)}.$$ The kernel function $K(\cdot)$ and bandwidth $h$ used here
may be different from those used in the N-W estimator, but for simplicity, we use the same notations to represent them.
Moreover, according to the constraint $\int_{\mathscr Y}f(y)dy=1,$ the standardized estimators of $f_\beta(y)$ and $f_\beta(Y_k)$ can be respectively expressed as
\begin{equation}\label{(3.8)}\widehat f_\beta(y)=\frac{\widetilde f_\beta(y)}{\widehat{\int_{\mathscr Y}\widetilde f_\beta(y)dy}},\ \widehat f_\beta(Y_k)=\frac{\widetilde f_\beta(Y_k)}{\widehat{\int_{\mathscr Y}\widetilde f_\beta(y)dy}},\end{equation} where $\widehat{\int_{\mathscr Y}\widetilde f_\beta(y)dy}$ is the estimator of $\int_{\mathscr Y}\widetilde f_\beta(y)dy$ defined by
$$\widehat{\int_{\mathscr Y}\widetilde f_\beta(y)dy}=\frac1n\sum_{i=1}^n\frac{\widetilde f_\beta(Y_i)}{\widehat p_Y(Y_i)}$$ with
$\widehat p_Y(Y_i)$ being the N-W estimator of the density $p_Y(Y_i)$ of $Y$ as
$$\widehat p_Y(Y_i)=\frac{1}{n-1}\sum_{j\neq i,j=1}^n\frac1h K((Y_j-Y_i)/h).$$

With the estimators $\widetilde f_\beta(y)$ and $\widetilde f_\beta(Y_k)$ or $\widehat f_\beta(y)$ and $\widehat f_\beta(Y_k)$ of the least favorable curve,
we get the estimated density as
\begin{equation}\label{(3.9)}\widehat p(y;\beta|{x})=\exp\left\{\beta^T{x}\cdot y-b(\beta^T{x},f^*_\beta)+\log f^*_\beta(y)\right\},\end{equation} where $f^*_\beta$ is the estimator $\widetilde f_\beta(\cdot)$ or the standardized version $\widehat f_\beta(\cdot)$. As a result, the estimated likelihood function for $\beta$ is obtained as
\begin{equation}\label{(3.10)}\widehat l(\beta)=\prod_{i=1}^n\exp\left\{\beta^T X_i\cdot Y_i-b(\beta^T X_i,f^*_\beta)+\log f^*_\beta(Y_i)\right\}.\end{equation} Therefore, the profile likelihood estimator of $\beta$ is obtained as
\begin{equation}\label{(3.11)}\widehat\beta=\arg\max_{\beta}\widehat l(\beta).\end{equation}
Finally, we get the estimators of $f(y)$ as
\begin{equation}\label{(3.12)}\widetilde f_{\widehat\beta}(y) \ \mbox{ and } \ \widehat f_{\widehat\beta}(y)=\frac{\widetilde f_{\widehat\beta}(y)}{\frac1n\sum_{i=1}^n\frac{\widetilde f_{\widehat\beta}(Y_i)}{\widehat p_Y(Y_i)}}.\end{equation}

The estimation procedure aforementioned is actually based on the profile likelihood for semiparametric model. The special feature above is that the least favorable curve obtains an explicit expression, and  consequently, its nonparametric estimators in (\ref{(3.6)}) and (\ref{(3.7)}) have explicit expressions. Thus, our methodology can be implemented and is computationally simple. Moreover, by semiparametric model theory (see, e.g., Severini and Wang (1992), and Bickel et al. (1993)), the resulting estimation can achieve semiparametric efficiency; for the details see Theorem 1 given below. On the other hand, the new likelihood function $\widehat l(\beta)$ in (\ref{(3.10)}) can be used for other statistical inferences such as the constructions of confidence interval and hypothesis test.

\subsection{Estimation for semiparametric outcome regression}
We now revisit the motivating example given in Section 2. For semiparametric exponential family distribution (\ref{(1.5)}), a least favorable curve can be chosen as
\begin{equation*}f^m_\beta(y)=\left(E\left(\frac{\exp\{\beta^T X\cdot y\}}{ \exp\{\int_{0}^{\beta^TX}E(Y|\beta^T x,\delta=1)d(\beta^T x)\}}\Big|y,\delta=1\right)\right)^{-1},\end{equation*} and the estimators of $f^m_\beta(y)$ and $f^m_\beta(Y_k)$ can be expressed as
\begin{eqnarray*}&&\widetilde f^m_\beta(y)=\left(\sum_{i\in \mathscr O}\frac{\exp\{\beta^T X_i\cdot y\}}{ \exp\left\{\int_0^{\beta^TX_i}\frac{\sum_{j\in \mathscr O} Y_jK((\beta^T X_j-t)/h)}{\sum_{j\in \mathscr O} K((\beta^T X_j-t)/h)}dt\right\}}W_i(y)\right)^{-1},\\&&
\widetilde f^m_\beta(Y_k)=\left(\sum_{i\neq k,i\in \mathscr O}\frac{\exp\{\beta^T X_i\cdot Y_k\}}{ \exp\left\{\int_0^{\beta^TX_i}\frac{\sum_{j\in \mathscr O} Y_jK((\beta^T X_j-t)/h)}{\sum_{j\in \mathscr O} K((\beta^T X_j-t)/h)}dt\right\}}W_i(Y_k)\right)^{-1},\end{eqnarray*} where $\mathscr O=\{i:\delta_i=1\}$, the index set of the observed data. By the relation $f^m(y)=\frac{1}{\sqrt{2\pi}}\exp(-\frac{y^2}{2})g_1(y)$, we obtain the estimators of $g_1(y)$ and $g_1(Y_k)$ respectively as
$$\widetilde g_{1\beta}(y)=\sqrt{2\pi}\widetilde f^m_\beta(y)\exp\left(\frac{y^2}{2}\right), \ \tilde g_{1\beta}(Y_k)=\sqrt{2\pi}\widetilde f^m_\beta(Y_k)\exp\left(\frac{Y_k^2}{2}\right).$$ By distribution (\ref{(1.5)}) and the relation given in (\ref{(1.6)}), we use the log-likelihood to estimate $\beta$ as
$$\widehat\beta_{\mathscr O}=\arg\min_{\beta}\sum_{i\in\mathscr O}\left\{\beta^TX_i\cdot Y_i-\frac12(\beta^TX_i)^2-G(\beta^TX_i,\widetilde g_{1\beta})+\log \widetilde f_\beta^m(Y_i)\right\},$$ where $$G(\beta^TX_i,\widetilde g_{1\beta})=\log\int_{\mathscr Y}\frac{1}{\sqrt{2\pi}}\exp\left\{-\frac{(y-\beta^TX_i)^2}{2}\right\}\widetilde g_{1\beta}(y)dy.$$ Consequently, the function $g(\beta^TX_i,g_1)$ in regression (\ref{(1.7)}) can be estimated by
$$g(\widehat\beta^T_{\mathscr O}X_i,\widetilde g_{1\widehat\beta_{\mathscr O}})=\frac{\int_{\mathscr Y}(y-\widehat\beta^T_{\mathscr O}X_i)\exp\left\{-\frac{(y-\widehat\beta^T_{\mathscr O}X_i)^2}{2}\right\}\widetilde g_{1\widehat\beta_{\mathscr O}}(y)dy}{\int_{\mathscr Y}\exp\left\{-\frac{(y-\widehat\beta^T_{\mathscr O}X_i)^2}{2}\right\}\widetilde g_{1\widehat\beta_{\mathscr O}}(y)dy}.$$ Finally,
for a new observed data $X$, we get the empirical version of the outcome regression (\ref{(1.7)}) as
\begin{equation}\label{(3.13)}\widehat E(Y|X)=\widehat\beta^T_{\mathscr O}X+g(\widehat\beta^T_{\mathscr O}X,\widetilde g_{1\widehat\beta_{\mathscr O}}).\end{equation}

Note that the maximum likelihood estimation above is different from the least squares estimation for regression (\ref{(1.7)}) obtained by minimizing
$$\sum_{i\in\mathscr O}(Y_i-\beta^TX_i- g(\beta^TX_i,\widetilde g_{1\beta}))^2,$$ because the maximum likelihood uses full information of the distribution and then results in an efficient estimation.

\subsection{Theoretical properties}

As shown above, we only consider the case when $y$ is a continuous variable and $f(y)$ is a smooth function of $y$ although the method can be extended to the case with discrete variable.
We then introduce following regularity conditions.
\begin{itemize}\item[C1.]
The density function $p_X(x)$ of $X$ has the second-order continuous and bounded derivative, and satisfies $p_X(x)>0$ for all $x$. The function $f=f(y)$ has the second-order continuous derivative on $\mathscr Y$.
\item[C2.] The expectation function $m_\beta(y)=E\left(\frac{\exp\{\beta^T X\cdot y\}}{ \int_{\mathscr Y}\exp\{\beta^T X\cdot y\}f(y)dy}\Big|y\right)$ has the second-order continuous derivative on $\mathscr Y$, and there exists the variance function $v^2_\beta(y)=Var\left(\frac{\exp\{\beta^T X\cdot y\}}{ \int_{\mathscr Y}\exp\{\beta^T X\cdot y\}f(y)dy}\Big|y\right)$.
\item[C3.] Parameter space $\mathscr B$ of $\beta$ is a closed bounded set.
\item[C4.]
Kernel function $K(u)$ is symmetric with
respect to $u=0$, and satisfies $\int K(u)du=1$, $\int u^2
K(u)du<\infty$ and $\int u^2 K^2(u)du<\infty$.
\item[C5.] The bandwidth satisfies $h\rightarrow 0$ and $nh\rightarrow\infty$.
\end{itemize}

The conditions C1-C5 are common obviously.
Denote $l(\beta,f)=\log p(y;\beta,f|{x})$ and let $\beta^0$ be the true value of $\beta$.
We have the following theorem.

\begin{thm}  Under the conditions C1-C5, the profile likelihood estimator $\widehat\beta$ defined in (\ref{(3.11)}) satisfies
$$\sqrt n(\widehat \beta-\beta^0)\stackrel{\mathscr D}\longrightarrow N(0,{\widehat i_{\beta^0}}^{-1}),$$ where $\stackrel{\mathscr D}\longrightarrow$ stands for convergence in distribution, and
$$\widehat i_\beta=E\left(\frac{\partial l(\beta,f)}{\partial\beta}\frac{\partial l(\beta,f)}{\partial\beta^T}\right)-\left(E\left(\frac{\partial l(\beta,f)}{\partial\beta}\frac{\partial l(\beta,f)}{\partial f}\right)\right)^2\left(E\left(\frac{\partial l(\beta,f)}{\partial f}\right)^2\right)^{-1}.$$
\end{thm}

The proof of the theorem is given in Appendix.
For the theorem and its proof, we have the following explanations:

{Remark 1.} {\it
\begin{itemize} \item[(1)] The theorem ensures that our estimator $\widehat \beta$ is semiparametric efficient because $\widehat i_\beta$ is the marginal Fisher information matrix for $\beta$ (Severini and Wang (1992)).
\item[(2)] Our proof is based mainly on the observation that our estimator $\widetilde f_\beta(y)$ in (\ref{(3.6)}) could be thought of as a local maximum likelihood estimation. It can be seen from the definition (\ref{(1.2)}) that for given $\beta$, the local likelihood function for $f$ can be expressed as
\begin{equation*}\prod_{i=1}^n\left(\beta^T X_i\cdot y-b(\beta^T X_i,f)+\log f(y)\right)K\left(\frac{Y_i-y}{h}\right).\end{equation*} By the local likelihood above, the derivative (\ref{(3.1)}) and the relation (\ref{(3.4)}), we get the local maximum likelihood estimator of $f(y)$ as $\widetilde f_\beta(y)$, which is the same as our estimator given in (\ref{(3.6)}). Then, by Lemma 2.4 of Severini and Wang (1992), such a local maximum likelihood estimator $\widetilde f_\beta(y)$ is a consistent estimator of a least favorable curve. Finally, with the estimator $\widetilde f_\beta(y)$, Proposition 2 of Severini and Wang (1992) shows that the resulting maximum likelihood estimator $\widehat\beta$ for $\beta$ satisfies the asymptotical normality as in Theorem 1.
\end{itemize}}

Denote by $p_{\beta}(x|y)$ the conditional density function of $\frac{\exp\{\beta^T X\cdot y\}}{ \int_{\mathscr Y}\exp\{\beta^T X\cdot y\}f(y)dy}$ for given $y$.
Let $V_{\beta}(y)=\frac{v^2_\beta(y)}{p_{\beta}(x|y)}\|K\|_2^2$, where $\|K\|_2^2=\int k^2(u)du$ and $v^2_\beta(y)$ is defined in C2.
For the nonparametric estimator $\widetilde f_{\widehat\beta}(y)$ of $f(y)$, we have the following theorem.

\begin{thm}
In addition to the conditions C1-C5, if $h=O(n^{-\delta})$ for some constant $0<\delta<1/5$, then,
$$\sqrt{nh}\left(\widetilde f_{\widehat\beta}(y)-f(y)\right)\stackrel{\mathscr D}\longrightarrow N\left(0,\frac{V_{\beta^0}(y)}{m^2_{\beta^0}(y)}\right),$$ where $m_{\beta}(y)$ is defined in C2.
\end{thm}

The proof of theorem is given in Appendix as well. The theorem guarantees the consistency and asymptotical normality of the nonparametric estimator $\widetilde f_{\widehat\beta}(y)$ of $f(y)$. The asymptotical variance of the estimator depends only on the conditional distribution of $\exp\{\beta^T X\cdot y-b(\beta^TX,f)\}$ for given $y$.

\setcounter{equation}{0}
\section{Simulation studies}

In this section, we use simulation study to assess the finite sample performance of our proposal, and compare ours with the marginal likelihood proposed by Ning et al. (2017). Because their marginal likelihood (\ref{(1.9)}) is computationally intensive, as shown in Introduction, the $2$th order information form given in (\ref{(1.11)}) is used as its surrogate.
For comprehensive comparison and assessment, several univariate regressions and multivariate regressions with complete or incomplete data are considered simultaneously.
The estimation mean, median, bias, mean squared error (MSE) and standard deviation (sd) of the parameter estimators are reported to evaluate the performances of the parametric estimation, and the estimation curve of the median of the estimators of the nonparametric function is used to examine the behavior of the nonparametric estimation. Throughout the simulation procedure, all the numerical results are obtained over 100 replications.

Our main goal is to check if our profile likelihood can obtain a high estimation efficiency and reduce the estimation bias as claimed in the previous sections.

{\it Experiment 1}. We begin with the following univariate linear regression model:
$$Y_i=\beta X_i+\epsilon_i, i=1,\cdots,n.$$ In the simulation procedure, the parameter of interesting $\beta$ is chosen as $\beta=2$, the variables $X_i,i=1,\cdots,n$, are independently generated from $N(\mu,1)$, and the error $\epsilon_i,i=1,\cdots,n$, are independently generated from $N(0,\sigma^2)$. Simulation results with different choices of $n$, $\mu$ and $\sigma^2$ are reported in Table 1 and Table 2, in which  $\widehat\beta$ and $\widehat\beta_N$ denote the parameter estimators of ours and the $2$th order information estimator of Ning et al. (2017), respectively.

Table 1 presents the simulation results for $n=100$.
From the Table 1, we have the following findings:
\begin{itemize}

\item[(1)] For the case of error variance $\sigma^2=1$,
our method is much better than that of Ning et al. (2017) since the MSE and standard deviation of our method are significantly smaller than those of Ning et al. (2017) uniformly for all the choices of $\mu$.
But the differences between the biases, means and medians of the two types of the estimators are not significant.
\item[(2)] For the case with $\mu=1$, our method is clearly superior to that of Ning et al. (2017) in the sense that the MSE and bias of ours is significantly smaller than those of Ning et al. (2017) uniformly for all the choices of $\sigma^2$. In this case, the estimators of Ning et al. (2017) have a non-negligible bias, and the mean and median of the estimators of Ning et al. (2017) are far away from the true value of the parameter. Particularly, when $\sigma^2=0.1$, the estimator $\widehat\beta_N$ of Ning et al. (2017) has a huge bias and a relatively large variance, even the estimator is corrupted. This just illustrates the second point of view given in Subsection 1.2.2: the conditional likelihood will result in a large bias of modeling when $\sigma^2$ is small.
\item[(3)] For the case of variance of error $\sigma^2=1.15$, also our method is much better then that of Ning et al. (2017) in the sense that the MSE of ours is significantly smaller than that of Ning et al. (2017) uniformly for all the choices of $\mu$. In this case, also the estimators of Ning et al. (2017) have a non-negligible bias, and the mean and median of the estimators of Ning et al. (2017) are far away from the true value of the parameter.
\item[(4)] In the case of large error variance, the variance of our estimator is slightly increased, and the difference between the estimation variances of the two types of estimators is not significant.
\end{itemize}

\begin{table}
\centering
\caption{Simulation results for Experiment 1 with $n=100$}\label{parset}
\begin{tabular}{c|c|ccccccc}
\hline
            &&mean&median&MSE&bias &sd  \\\hline
            &&&&$\sigma^2$=1\\
$\mu=1$&$\widehat\beta$&2.1081&2.0673&0.0896&0.1081&0.2791 \\
 &$\widehat\beta_{N}$&2.0665&2.0693&0.0999&0.0665&0.3091\\
$\mu=2$&$\widehat\beta$&2.0525&2.0097&0.0525&0.0525&0.2232 \\
 &$\widehat\beta_{N}$&2.0724&2.0232&0.1607&0.0724&0.3943\\
$\mu=3$&$\widehat\beta$&2.0149&1.9967&0.0335&0.0149&0.1826 \\
 &$\widehat\beta_{N}$&2.0584&1.9924&0.1171&0.0584&0.3371\\
           &&&& $\mu$=1\\
$\sigma^2=0.1$&$\widehat\beta$&2.3332&2.2980&0.1253&0.3332&0.1195 \\
&$\widehat\beta_{N}$&211.7805&205.9996&45515.95&209.7805&38.8343\\
$\sigma^2=1.1$&$\widehat\beta$&2.0588&2.0241&0.0773&0.0588&0.2718 \\
 &$\widehat\beta_{N}$&1.7353&1.6885&0.1543&0.2647&0.2903\\
$\sigma^2=1.15$&$\widehat\beta$&2.0121&1.9958&0.0871&0.0121&0.2948 \\
&$\widehat\beta_{N}$&1.5269&1.5123&0.2941&0.4731&0.2652\\
         &&&& $\sigma^2$=1.15\\
$\mu=0$&$\widehat\beta$&2.0458&2.0207&0.1090&0.0458&0.3270 \\
&$\widehat\beta_{N}$&1.5457&1.5210&0.2757&0.4542&0.2636\\
$\mu=1$&$\widehat\beta$&2.0503&2.0117&0.0825&0.0503&0.2829 \\
&$\widehat\beta_{N}$&1.6169&1.5769&0.2314&0.3830&0.2910\\
$\mu=2$&$\widehat\beta$&2.0245&2.0046&0.0712&0.0245&0.2658 \\
&$\widehat\beta_{N}$&1.5778&1.5761&0.2535&0.4221&0.2744\\
$\mu=3$&$\widehat\beta$&2.0144&1.9994&0.0499&0.0144&0.2231 \\
&$\widehat\beta_{N}$&1.5590&1.5427&0.2495&0.4409&0.2347\\
\hline
\end{tabular}
\end{table}

Table 2 gives the simulation results for $n=200$ and 400. Similar to the findings from Table 1 with $n=100$, for the cases of $n=200$ and 400,
our method is also much better than that of Ning et al. (2017) in the sense that ours can significantly reduce MSE and bias. Particularly, when $\sigma^2=0.1$, the estimator $\widehat\beta_N$ of Ning et al. (2017) has a huge bias and a relatively large variance, which again illustrates the point of view: the conditional likelihood will result in a large bias of modeling when $\sigma^2$ is small. Moreover, by comparing the results in Table 2 with those in Table 1, we can see
the MSE of our estimator $\widehat\beta$ tends to zero with a fast speed as the sample size $n$ increases, implying the consistency of the estimator $\widehat\beta$. But the convergence rate of the estimator $\widehat\beta_N$ is relatively slow. Similar to the case of $n=100$, when the error variance is large, the estimation variance of ours is slightly increased, and sometimes is larger than those of the estimator of Ning et al. (2017).

\begin{table}
\centering
\caption{Simulation results for Experiment 1 with $n=200$ and 400}\label{parset}
\begin{tabular}{c|c|ccccccc}
\hline
                   &&mean&median&MSE&bias &sd  \\\hline
                   &&&&$n=200$, $\mu$=1\\
$\sigma^2=0.1$&$\widehat\beta$&2.3202&2.2920&0.1172&0.3202&0.1210 \\
&$\widehat\beta_{N}$&203.4608&202.3437&41156.9&201.4608&23.8837\\
$\sigma^2=1.1$&$\widehat\beta$&2.0913&2.0439&0.0665&0.0913&0.2412 \\
 &$\widehat\beta_{N}$&1.6687&1.6645&0.1427&0.3312&0.1816\\
$\sigma^2=1.15$&$\widehat\beta$&2.0672&2.0435&0.0664&0.0672&0.2489 \\
&$\widehat\beta_{N}$&1.5615&1.5439&0.2196&0.4385&0.1655\\
         &&&&$n=200$, $\sigma^2$=1.15\\
$\mu=0$&$\widehat\beta$&2.0875&2.0470&0.1134&0.0875&0.3253 \\
&$\widehat\beta_{N}$&1.5198&1.5082&0.2600&0.4801&0.1716\\
$\mu=2$&$\widehat\beta$&2.0240&1.9910&0.0410&0.0240&0.2011 \\
&$\widehat\beta_{N}$&1.5553&1.5473&0.2296&0.4446&0.1785\\
$\mu=3$&$\widehat\beta$&2.0304&2.0032&0.0295&0.0304&0.1691 \\
&$\widehat\beta_{N}$&1.5666&1.5662&0.2133&0.4334&0.1597\\
         &&&&$n=400$, $\mu$=1\\
$\sigma^2=0.1$&$\widehat\beta$&2.314&2.2805&0.1323&0.3146&0.1824 \\
 &$\widehat\beta_{N}$&201.1039&203.3637&40296.92&200.1039&15.9798\\
$\sigma^2=1.1$&$\widehat\beta$&2.0471&2.0116&0.0333&0.0471&0.1763 \\
 &$\widehat\beta_{N}$&1.6889&1.6793&0.1167&0.3110&0.1413\\
$\sigma^2=1.15$&$\widehat\beta$&2.0372&2.0003&0.0425&0.0372&0.2028 \\
&$\widehat\beta_{N}$&1.5490&1.5361&0.2229&0.4509&0.1399\\
       &&&&$n=400$, $\sigma^2$=1.15\\
$\mu=0$&$\widehat\beta$&2.0665&1.9927&0.0724&0.0665&0.2608 \\
&$\widehat\beta_{N}$&1.5105&1.5116&0.2547&0.4894&0.1231\\
$\mu=2$&$\widehat\beta$&2.0109&2.0041&0.0169&0.0109&0.1298 \\
&$\widehat\beta_{N}$&1.5190&1.5033&0.2489&0.4809&0.1328\\
$\mu=3$&$\widehat\beta$&2.0008&1.9968&0.0186&0.0008&0.1365 \\
&$\widehat\beta_{N}$&1.5067&1.4927&0.2567&0.4932&0.1161\\
\hline
\end{tabular}
\end{table}

It is worth noting, on the other hand, that our new method can consistently estimate the unknown base measure function $f(y)$. As shown above, the true base measure function is $f(y)=\exp(-y^2/2)$.
Under the situation of $\mu=0$ and $\sigma^2=1.15$, the median (solid line) of estimators $\widehat f(y)$ and the true function $f(y)$ (dot line) are presented in Figure 2. It can be seen that the estimated function $\widehat f(y)$ is very close to the true one, implying the consistency of the estimator.

\begin{figure}[H]
\caption[]{The median (solid line) of the estimated curves of $f(y)$ and the true curve (dot line) of the function $f(y)=\exp(-y^2/2)$ for Experiment 1 with $\mu=0$ and $\sigma^2=1.15$.}
\label{penG}
\includegraphics[height=7cm]{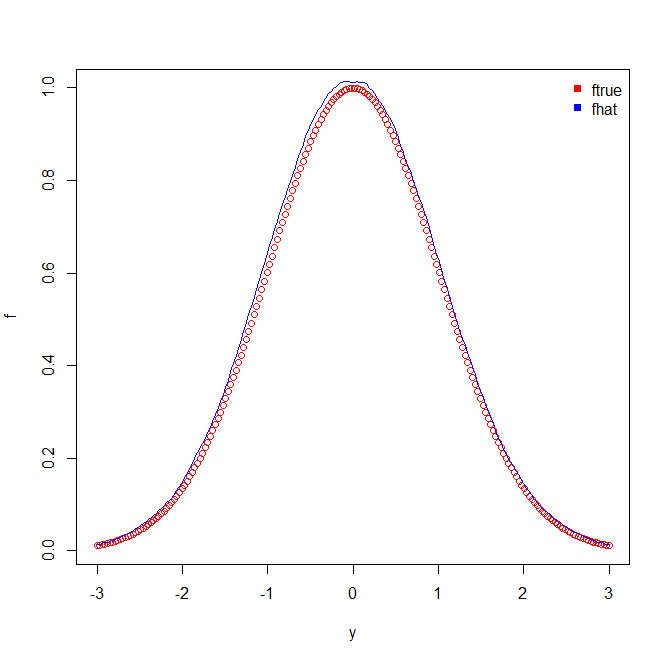}
\end{figure}

{\it Experiment 2}. We next consider the multivariate linear regression model: $$Y_i=\beta_1X_{1i}+\beta_2X_{2i}+\beta_3X_{3i}+\epsilon_i, i=1,\cdots ,n,$$
where the covariate vector $X_i=(X_{1i},X_{2i},X_{2i})^T, i=1,\cdots ,n,$ are independently generated from $N(0,{\Sigma})$, where $\Sigma=(\sigma_{ij})$ is the covariance matrix with $\sigma_{ij}=0.1^{|i-j|}$, and the errors $\epsilon_i, i=1,\cdots ,n$, are independent and identically distributed as $N(0, \sigma^2)$, and are independent of $X_i$. We set $\beta_1=1$, $\beta_2=2$ and $\beta_3=3$ in the procedure of simulation. The simulation results is reported in Table 3. For the multivariate linear regression, our estimators $\widehat\beta_1$, $\widehat\beta_2$ and $\widehat\beta_3$ are usually superior to those of Ning et al. (2017). Furthermore, it is seen that the bias and MSE of the parameter estimators of Ning et al. (2017) are very large when the error variance is smaller than 1. On the other hand, as the sample size $n$ increases, the estimation mean of $\beta=(\beta_1, \beta_2, \beta_3)^T$ obtained by our new method gets closer to the true value $\beta=(1,2,3)^T$, and the MSE tends to zero.

Moreover, for such a distribution, the true base measure function is $f(y)=\exp(-y^2/2)$. The median of the estimators of the base measure function $f(y)$ is given in Figure 3, which again demonstrates the consistency of the nonparametric estimator.

\begin{table}
\centering
\caption{Simulation results for Experiment 2}\label{parset}
\begin{tabular}{c|c|ccccccc}
\hline
         &Criteria&$\beta_1$&$\beta_2$&$\beta_3$&$\beta_{1N}$ &$\beta_{2N}$&$\beta_{3N}$  \\\hline
             &&&&$n=100$&&\\
$\sigma^2=0.1$&mean&1.0664&2.1482&3.2522&110.3418&220.9349&331.2640 \\
              &MSE&0.0424&0.0461&0.0875&12740.17&51086.38&114837.26\\
              &Bias&0.0664&0.1482&0.2522&109.3418&218.9349&328.2640\\
$\sigma^2=0.5$&mean&1.0953&2.0771&3.1956&4.1825&8.3021&12.4636\\
              &MSE&0.0407&0.0523&0.0792&10.5943&41.5347&93.2844\\
              &Bias&0.0953&0.0771&0.1956&3.1835&6.3021&9.4637\\
$\sigma^2=1.0$&mean&1.0210&1.9190&2.9740&1.0536&2.1662&3.2024 \\
              &MSE&0.0535&0.0737&0.0473&0.0407&0.1778&0.3580\\
              &Bias&0.0210&0.0809&0.0259&0.0536&0.1662&0.2024\\
               &&&&$n=200$&&\\
$\sigma^2=0.1$&mean&1.0760&2.1558&3.2467&103.1370&206.1927&309.5052 \\
              &MSE&0.0127&0.0312&0.0659&10737.79&42899.99&96663.79\\
              &Bias&0.0760&0.1558&0.2467&102.13708&204.1927&306.5052\\
$\sigma^2=0.5$&mean&1.0699&2.0669&3.1906&4.1641&8.3045&12.4247\\
              &MSE&0.0375&0.0533&0.0649&10.3057&40.7452&91.0900\\
              &Bias&0.0699&0.0699&0.1906&3.1641&6.3045&9.4247\\
$\sigma^2=1.0$&mean&1.0062&1.9357&3.0480&1.0361&2.1017&3.1121 \\
              &MSE&0.0370&0.0661&0.0507&0.0209&0.0713&0.1437\\
              &Bias&0.0062&0.0642&0.0480&0.0361&0.1017&0.1121\\
\hline
\end{tabular}
\end{table}

\begin{figure}[H]
\caption[]{ The median curve (solid line) of the estimators of $f(y)$ and the true curve (dot line) of the function $f(y)=\exp(-y^2/2)$ in Experiment 2 with the variance error $\sigma^2=1$}
\label{penG}
\includegraphics[height=7cm]{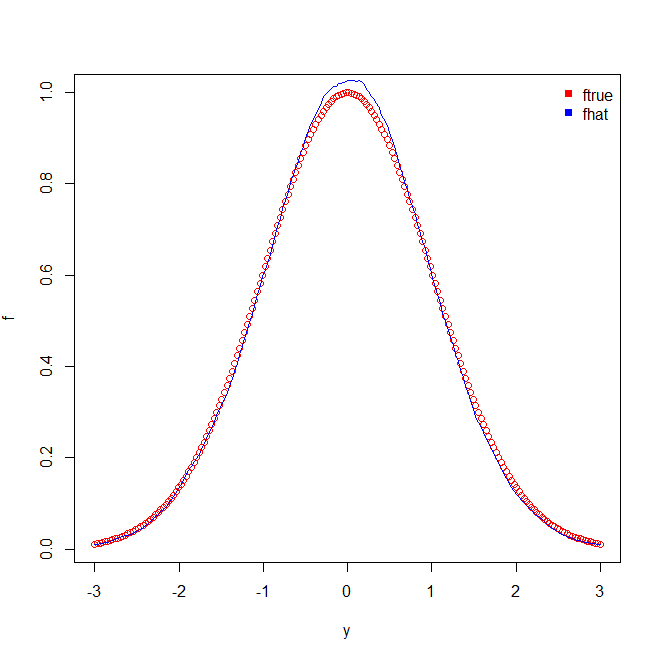}
\end{figure}

{\it Experiment 3.} We now consider the motivating example with missing data given in Section 2. Suppose
$$Y_i=\beta X_i+\epsilon_i, i=1,\cdots,n,$$
where the parameter $\beta=2$, the variables $X_i,i=1,\cdots,n,$ are independent and identically distributed as $N(2,1)$, and the errors $\epsilon_i,i=1,\cdots,n,$ are independent and follow the normal distribution $N(0,1.1)$, and are independent of $X_i$. The following two missing mechanisms are considered to create missing values:
\begin{itemize}\item[(1)]
$P(\delta_i=1|X_i,Y_i)=c\,I\{X_i> 0\}I\{Y_i> 0\}$,
\item[(2)]
$P(\delta_i=1|X_i,Y_i)=c\,I\{1.8X_i< Y_i\}$,
\end{itemize}
where the constant $0\leq c\leq1$ and $I\{\cdot\}$ is an indicator function. The first missing mechanism (1) means that $X_i$ and $Y_i$ are observed with probability $c$ if $X_i>0$ and $Y_i>0$, and $X_i$ and $Y_i$ are missing with probability $1$ if $X_i\le0$ or $Y_i\le0$. This missing mechanism satisfies the decomposable condition as defined in Subsection 2.1. Thus the conditional distribution of $Y$ given $X=x$ and $\delta=1$ belongs to the semiparametric exponential family. The second missing mechanism (2) means that $X_i$ and $Y_i$ are observed with probability $c$ if $1.8X_i<Y_i$, and $X_i$ and $Y_i$ are missing with probability $1$ if $1.8X_i\geq Y_i$. But the second missing mechanism (2) does not satisfy the decomposable condition.
In the following, we use the observed $X_i$ and $Y_i$ (i.e., the data with $\delta_i=1$), together with the methodologies of ours and Ning et al. (2017), to estimate $\beta$.

Under the first missing mechanism (1), the observation probability $c$ is chosen as  0.6, 0.7 and 0.8 in the simulation study, and the simulation results are listed in Table 4. With the second missing mechanism (2), the observation probability $c$ is chosen as 0.85, 0.90 and
0.95, and the simulation results are reported in Table 5.
We see that for both cases the estimator $\widehat\beta$ of our proposal is
significantly superior to the estimator $\widehat\beta_N$ of Ning et al. (2017) under all the criteria and for all the choices of observation probability $c$. Moreover, our method is robust to the choice of the missing mechanism, i.e., whether the decomposable condition holds or not, our method always obtains a better estimator of $\beta$. However, without the decomposable condition, the conditional likelihood results in more bad estimation behavior.

\begin{table}
\centering
\caption{Simulation results for Experiment 3 with missing mechanism (1)}\label{parset}
\begin{tabular}{c|c|ccccccc}
\hline
           missing prob. &&mean&median&MSE&bias &sd  \\\hline
           &&&&$n=100$&&\\
$c=0.6$&$\widehat\beta$&2.0290&1.9839&0.0976&0.0290&0.3111 \\
 &$\widehat\beta_{N}$&1.7508&1.6681&0.2899&0.2491&0.4774\\
$c=0.7$&$\widehat\beta$&2.0299&2.0177&0.0870&0.0299&0.2934 \\
&$\widehat\beta_{N}$&1.7264&1.6768&0.2052&0.2735&0.3610\\
$c=0.8$&$\widehat\beta$&2.0340&2.0268&0.0624&0.0340&0.2475 \\
&$\widehat\beta_{N}$&1.7474&1.7422&0.1535&0.2525&0.2996\\
  &&&&$n=200$&&\\
$c=0.6$&$\widehat\beta$&2.0550&2.0129&0.0667&0.0550&0.2523 \\
&$\widehat\beta_{N}$&1.7580&1.7413&0.1414&0.2420&0.2879\\
$c=0.7$&$\widehat\beta$&2.0301&1.9935&0.0455&0.0301&0.2112 \\
&$\widehat\beta_{N}$&1.7276&1.7037&0.1668&0.2724&0.3043\\
$c=0.8$&$\widehat\beta$&2.0124&1.9792&0.0416&0.0124&0.2036 \\
&$\widehat\beta_{N}$&1.6832&1.6798&0.1540&0.3168&0.2316\\
  &&&&$n=400$&&\\
$c=0.6$&$\widehat\beta$&2.0169&1.9844&0.0257&0.0169&0.1595 \\
 &$\widehat\beta_{N}$&1.6693&1.6489&0.1442&0.3307&0.1866\\
$c=0.7$&$\widehat\beta$&2.0208&1.9845&0.0212&0.0208&0.1442 \\
 &$\widehat\beta_{N}$&1.6427&1.6354&0.1574&0.3573&0.1727\\
$c=0.8$&$\widehat\beta$&2.0545&2.0418&0.0242&0.0545&0.1457\\
&$\widehat\beta_{N}$&1.6708&1.6725&0.1314&0.3292&0.1518\\
\hline
\end{tabular}
\end{table}

\begin{table}
\centering
\caption{Simulation results for Example 3 with missing mechanism (2)}\label{parset}
\begin{tabular}{c|c|ccccccc}
\hline
           missing prob. &&mean&median&MSE&bias &sd  \\\hline
           &&&&$n=100$&&\\
$c=0.85$&$\widehat\beta$&2.0262&1.9778&0.0762&0.0262&0.2747 \\
 &$\widehat\beta_{N}$&3.6683&3.3660&3.7960&1.6682&1.0064\\
$c=0.90$&$\widehat\beta$&2.0283&2.0146&0.0607&0.0283&0.2448\\
&$\widehat\beta_{N}$&3.6829&3.5448&3.9384&1.6829&1.0517 \\
$c=0.95$&$\widehat\beta$&1.9707&1.9650&0.0581&0.0292&0.2393 \\
&$\widehat\beta_{N}$&3.7520&3.7129&3.9043&1.7520&0.9136\\
  &&&&$n=200$&&\\
$c=0.85$&$\widehat\beta$&1.9925&1.9697&0.0370&0.0075&0.1923 \\
&$\widehat\beta_{N}$&3.4856&3.4072&2.6030&1.4856&0.6293\\
$c=0.90$&$\widehat\beta$&2.0061&1.9804&0.0339&0.0061&0.1841\\
&$\widehat\beta_{N}$&3.5679&3.5308&2.8701&0.5679&0.6416\\
$c=0.95$&$\widehat\beta$&1.9705&1.9620&0.0277&0.0295&0.1639 \\
&$\widehat\beta_{N}$&3.5244&3.3907&2.6766&1.5244&0.5938\\
 &&&&$n=400$&&\\
$c=0.85$&$\widehat\beta$&2.0041&1.9629&0.0331&0.0041&0.1820 \\
&$\widehat\beta_{N}$&3.4569&3.2750&2.3711&1.4569&0.4986\\
$c=0.90$&$\widehat\beta$&1.9869&1.9633&0.0245&0.0131&0.1561 \\
 &$\widehat\beta_{N}$&3.4176&3.4140&2.1454&1.4176&0.3684\\
$c=0.95$&$\widehat\beta$&1.9786&1.9659&0.0160&0.0214&0.1250 \\
&$\widehat\beta_{N}$&3.5216&3.4719&2.4887&1.5216&0.4163\\
\hline
\end{tabular}
\end{table}

In summary, all the simulation studies can illustrate that our profile likelihood estimation for the semiparametric exponential family is consistent and efficient, and is robust to different model conditions and specially to the different choices of the error variance, and is also robust to the choice of missing mechanism. Moreover, the profile likelihood is significantly superior to the existing methodology for most cases considered, specially for the case with small error variance and the case without the decomposable condition.

\appendix

\section{ Proofs of main results}

{\it Proof of (\ref{(1.3)})}. The first equation is a known property of the regularity distribution family. We only need to prove the second one.

Let $l(\theta,f)=\theta\cdot y-\log \int_{\mathscr Y}\exp\{\theta\cdot y\}f(y)dy+\log f(y).$ We first prove
\begin{equation}\label{(a1)}\frac{\partial l(\theta,f)}{\partial f}=-\frac{\exp\{\theta\cdot y\}}{\int_{\mathscr Y}\exp\{\theta\cdot y\}f(y)dy}+\frac{1}{f(y)}.\end{equation}
By the definition of the functional derivative, we have
\begin{eqnarray*}&&\hspace{-4ex}\frac{\delta(\log \int_{\mathscr Y}\exp\{\theta\cdot y\}f(y)dy)}{\delta f(z)}\\&&\hspace{-4ex}=\lim_{\varepsilon\rightarrow 0}\frac{\log \int_{\mathscr Y}\exp\{\theta\cdot y\}(f(y)+\varepsilon\delta(y-z))dy-\log \int_{\mathscr Y}\exp\{\theta\cdot y\}f(y)dy}{\varepsilon}
\\&&\hspace{-4ex}=\lim_{\varepsilon\rightarrow 0}\frac{\log (\int_{\mathscr Y}\exp\{\theta\cdot y\}f(y)dy+\varepsilon\int_{\mathscr Y}\exp\{\theta\cdot y\}\delta(y-z)dy)-\log \int_{\mathscr Y}\exp\{\theta\cdot y\}f(y)dy}{\varepsilon}
\\&&\hspace{-4ex}=\lim_{\varepsilon\rightarrow 0}\frac{\log (\int_{\mathscr Y}\exp\{\theta\cdot y\}f(y)dy+\varepsilon\exp\{\theta z\})-\log \int_{\mathscr Y}\exp\{\theta\cdot y\}f(y)dy}{\varepsilon}\left(\frac{0}{0}\right)\\&&\hspace{-4ex}=
\frac{\exp\{\theta z\}}{\int_{\mathscr Y}\exp\{\theta\cdot y\}f(y)dy}.\end{eqnarray*} Similarly, the derivative of $\log f(y)$ is $\frac{1}{f(y)}$. Then, the two results above implies (\ref{(a1)}).

Denote $p(y;\theta,f)=\exp\left\{\theta\cdot y-\log \int_{\mathscr Y}\exp\{\theta\cdot y\}f(y)dy+\log f(y)\right\}$. By the same argument as used above and (\ref{(a1)}), we can prove that the functional derivative of $p(y;\theta,f)$ is
\begin{eqnarray}\label{(a2)}\nonumber\frac{\partial p(y;\theta,f)}{\partial f}&=&\left(-\frac{\exp\{\theta\cdot y\}}{\int_{\mathscr Y}\exp\{\theta\cdot y\}f(y)dy}+\frac{1}{f(y)}\right)p(y;\theta,f)\\&=&\frac{\partial l(\theta,f)}{\partial f}p(y;\theta,f).\end{eqnarray}

Finally, we consider the case of $\theta(x)=\beta^Tx$. Not that $\int \frac{p(y;\theta(x),f|x)p_X(x)}{p_Y(y)}dx=1$ implies
$\frac{\partial}{\partial f}\int \frac{p(y;\theta(x),f|x)p_X(x)}{p_Y(y)}dx=0.$ Moreover, $\frac{\partial}{\partial f}\int\frac{p(y;\theta(x),f|x)p_X(x)}{p_Y(y)}dx=0$ is equivalent to
$\int \frac{\partial}{\partial f}\frac{p(y;\theta(x),f|x)p_X(x)}{p_Y(y)}dx=0.$ By combing this result with  (\ref{(a2)}), we have
\begin{eqnarray*}E\left(\frac{\partial l(\theta(x),f)}{\partial f}\Big|y\right)&=&\int \frac{\partial l(\theta(x),f)}{\partial f} \frac{p(y;\theta(x),f|x)p_X(x)}{p_Y(y)}dx\\&=&\int \frac{\partial}{\partial f}\frac{p(y;\theta(x),f|x)p_X(x)}{p_Y(y)}dx=0.\end{eqnarray*} The proof is completed.
$\square$

\

{\it Proof of (\ref{(3.1)})}. The proof follows directly from (\ref{(a1)}). $\square$

\

{\it Proof of Theorem 1.} By the profile likelihood theory for semiparametric model (see, e.g., Severini and Wang (1992), and Bickel et al. (1993)), we first prove that $\widetilde f_\beta(y)$ in (\ref{(3.6)}) is a consistent estimator of $f_\beta(y)$ given in (\ref{(3.5)}).

Based on the property of kernel estimator, we have that for given $\beta$ and any function $m(x)$ that has the second-order continuous derivative, the following asymptotic representations hold:
\begin{eqnarray*}&&\frac{\sum_{j=1}^n Y_jK(\beta^T(X_j-x)/h)}{\sum_{j=1}^n K(\beta^T(X_j-x)/h)}=E(Y|\beta^Tx)+o_p(1)
=b'(\beta^T x,f)+o_p(1),\\&&\frac1n\sum_{i=1}^nm(X_i)W_i(y)=
E(m(X)|Y=y)+o_p(1).\end{eqnarray*}
Then, for given $\beta$,
\begin{eqnarray*}\widetilde f_{\beta}(y)&=&\left(\sum_{i=1}^n\frac{\exp\{\beta^T X_i\cdot y\}}{ \exp\left\{\int_0^{\beta^T X_i}\frac{\sum_{j=1}^n Y_jK((\beta^T X_j-t)/h)}{\sum_{j=1}^n K((\beta^T X_j-t)/h)}dt\right\}}W_i(y)\right)^{-1}\\&=&\left(\sum_{i=1}^n\frac{\exp\{\beta^T X_i\cdot y\}}{ \exp\left\{\int_0^{\beta^TX_i}b'(\beta^T x,f)dx\right\}}W_i(y)\right)^{-1}+o_p(1)
\\&=&\left(\sum_{i=1}^n\frac{\exp\{\beta^T X_i\cdot y\}}{ \exp\left\{b(\beta^T X_i,f)\right\}}W_i(y)\right)^{-1}+o_p(1)\\&=&\left(\sum_{i=1}^n\frac{\exp\{\beta^T X_i\cdot y\}}{ \int_{\mathscr Y}\exp\{\beta^T X_i\cdot y\}f(y)dy}W_i(y)\right)^{-1}+o_p(1)\\&=&\left(E\left(\frac{\exp\{\beta^T X\cdot y\}}{ \int_{\mathscr Y}\exp\{\beta^T X\cdot y\}f(y)dy}\Big|y\right)\right)^{-1}+o_p(1).\end{eqnarray*} By the above and condition C3, we have
$$\sup_{\beta\in\mathscr B}\left(\widetilde f_{\beta}(y)-\left(E\left(\frac{\exp\{\beta^T X\cdot y\}}{ \int_{\mathscr Y}\exp\{\beta^T X\cdot y\}f(y)dy}\Big|y\right)\right)^{-1}\right)\stackrel{P}\longrightarrow 0.$$
This ensures that $\widetilde f_\beta(y)$ is a consistent estimator of $f_\beta(y)$.

We next prove that $\widetilde f_\beta(y)$ is a consistent estimator of a least favorable curve, i.e., the function $f_\beta(y)$ given in (\ref{(3.5)}) is a least favorable curve.
It can be seen from the derivative (\ref{(3.1)}) and the relation (\ref{(3.4)}) that the estimator $\widetilde f_\beta(y)$ in (\ref{(3.6)}) satisfies
\begin{eqnarray*}\widetilde f_\beta(y)=\argmin_{f}\prod_{i=1}^n\left(\beta^T X_i\cdot y-b(\beta^T X_i,f)+\log f(y)\right) K\left(\frac{Y_i-y}{h}\right).\end{eqnarray*}
By Lemma 2.4 of Severini and Wang (1992), the estimator $\widetilde f_\beta(y)$ is a consistent estimator of a least favorable curve, implying that actually $f_\beta(y)$ is a least favorable curve.

Finally, the above results, together with Proposition 2 of Severini and Wang (1992), yield that the resulting maximum likelihood estimator $\widehat\beta$ for $\beta$ satisfies the asymptotical normality as shown in Theorem 1.  $\square$

\

{\it Proof of Theorem 2.} From the Theorem 1 and its proof we can see that asymptotically,
$$\sum_{i=1}^n\frac{\exp\{\widehat\beta^T X_i\cdot y\}}{ \exp\left\{\int_0^{\widehat\beta^TX_i}\frac{\sum_{j=1}^n Y_jK((\widehat\beta^T X_j-t)/h)}{\sum_{j=1}^n K((\widehat\beta^T X_j-t)/h)}dt\right\}}W_i(y)$$ is identically distributed as
$$\sum_{i=1}^n\frac{\exp\{{\beta^0}^T X_i\cdot y\}}{ \int_{\mathscr Y}\exp\{{\beta^0}^T X_i\cdot y\}f(y)dy}W_i(y).$$ Then, by the asymptotical normality of kernel estimator, we have
$$\sqrt{nh}\left(\sum_{i=1}^n\frac{\exp\{{\beta^0}^T X_i\cdot y\}}{ \int_{\mathscr Y}\exp\{{\beta^0}^T X_i\cdot y\}f(y)dy}W_i(y)-m_{\beta^0}(y)\right)\stackrel{\mathscr D}\longrightarrow N(0,V_{\beta^0}(y)).$$
By the above result and Delta Method, we have
$$\sqrt{nh}\left(\widetilde f_{\widehat\beta}(y)-f(y)\right)\stackrel{\mathscr D}\longrightarrow N\left(0,\frac{V_{\beta^0}(y)}{m^2_{\beta^0}(y)}\right).$$

\end{document}